\documentstyle[emulateapj]{article}

\lefthead{Reunanen et al.}
\righthead{Near--infrared line imaging of NGC~7771}

\begin{document}

\submitted{Accepted for publication in ApJ}

\title{Near--infrared line imaging of the circumnuclear starburst ring\\ in NGC~
7771}

\author{J. Reunanen and J. K. Kotilainen}
\affil{Tuorla Observatory, University of Turku, V\"ais\"al\"antie 20, 
FIN--21500 Piikki\"o; reunanen@astro.utu.fi, jkotilai@stardust.astro.utu.fi}

\author{S. Laine}
\affil{Department of Physical Sciences, University of Hertfordshire, College 
Lane, Hatfield, Herts. AL10 9AB, U.K.; seppo@star.herts.ac.uk}

\and

\author{S. D. Ryder}
\affil{Joint Astronomy Centre, 660 N. A'Ohoku Place, Hilo, HI 96720; 
sryder@jach.hawaii.edu}

\begin{abstract}

We present high spatial resolution near--infrared broad-band $JHK$ 
images and, for the first time, Br$\gamma$ 2.1661~$\mu$m and H$_2$ 1--0 S(1) 
2.122~$\mu$m emission line images of the circumnuclear star forming ring 
(major axis diameter 7$''$ = 2 kpc) in the starburst galaxy NGC~7771. These 
data are used to investigate the morphology and extinction of the starburst 
ring and to study its star forming properties and history by comparing the 
observed quantities with an evolutionary population synthesis model. 

The clumpy morphology of NGC~7771 varies strongly with wavelength, due to the 
combination of extinction (for which we derive an average value of 
A$_V$ = 2.8), emission from hot dust and red supergiants, and several stellar 
generations in the ring. Also, the ellipticity and the position angle of the 
ring depend on the wavelength.

The starburst ring in NGC~7771 exhibits small Br$\gamma$ equivalent widths. 
Assuming a constant star formation model with M$_u$ = 100 M$_\odot$ results 
in very long lifetimes of the star forming regions (up to 1 Gyr), in 
disagreement with the clumpy near--infrared morphology and the observed radio 
spectral index of NGC~7771. This situation is only slightly remedied by 
assuming a reduced upper mass cutoff (M$_u$ = 30 M$_\odot$), resulting in ages
 between 8 and 180 Myr. We prefer an instantaneous star formation model with 
M$_u$ = 100 M$_\odot$ which can explain the derived Br$\gamma$ equivalent 
widths if a single starburst occurred 6--7 Myr ago. The main excitation 
mechanism of the molecular gas, based on the observed S(1)/Br$\gamma$ ratio, 
appears to be excitation by UV radiation from hot young stars. We derive 
M $\simeq$ 1900 M$_\odot$ for the mass of the excited H$_2$.

\end{abstract}


\section{Introduction}

NGC~7771 (UGC 12815) is a very luminous nearby (v$_{sys}$ = 4256 km 
s$^{-1}$) highly inclined ($\sim$69$^\circ$, Smith et al. 1999; hereafter 
S99) SBa galaxy, which forms a well-known pair with NGC~7770. At the distance 
of 56.7 Mpc (H$_0$ = 75 Mpc$^{-1}$ km s$^{-1}$) 1$''$ corresponds to 280 pc. 
For a large scale view of the pair, see e.g. S99 (their Fig. 1). The optical 
morphology of NGC~7771 is strongly affected by heavy extinction and prominent 
dust lanes in the bar. NGC~7771 is interacting with NGC~7770 and NGC~7771A, 
as indicated by the H$\alpha$ rotation curve (Keel 1993) and the HI velocity 
field (Nordgren et al. 1997).   

There is no sign of a Seyfert nucleus in NGC~7771. Most of the luminosity of 
NGC~7771 arises from a massive circumnuclear starburst, which was probably 
triggered by the interaction with NGC~7770. Nuclear star formation (SF) in 
NGC~7771 is concentrated in an elliptical ring with a major axis of 
$\sim$7$''$ (2 kpc), surrounding the starburst nucleus. This ring is clumpy 
and contains several emission regions both in radio (Neff \& Hutchings 1992) 
and near--infrared (NIR) continuum (S99). 

S99 noted an anticorrelation between the radio and NIR peaks, and explained 
this by two starburst episodes $\sim$10 Myr (radio regions) and $\sim$ 4--8 
Myr ago (NIR regions). A lower limit (4 Myr) for the age of the starburst has 
also been estimated from the dominant presence of red supergiants (RSG), 
based on NIR CO spectroscopy (Smith et al. 1996). The inner regions of 
NGC~7771 are also much more luminous in the $K$-band than those in normal 
galaxies (S99), which can be explained easily if there is a substantial 
population of RSGs in NGC~7771.

We present high resolution NIR $JHK$, and for the first time, Br$\gamma$ and 
H$_2$ emission line images of the circumnuclear ring in NGC~7771. Br$\gamma$ 
emission arises from HII regions surrounding young, hot OB stars, while H$_2$ 
1--0 S(1) emission (hereafter S(1)) arises from hot molecular gas and traces 
the material available for SF. The ratio between these two lines can give 
clues about the excitation mechanism(s) of the hot molecular gas (Puxley, 
Hawarden \& Mountain 1990). 

Our paper is organized in the following way. In Section 2 we discuss the 
observations and data reduction. In Sections 3 and 4 we discuss 
the evolutionary model used and morphology of the circumnuclear ring and 
constrain its star forming properties, stellar populations and star forming 
history by a detailed comparison among the NIR tracers and also compare them 
with the radio emission. Conclusions are presented in Section 5. 

\section{Observations and data reduction}

Observations were made in September 1998 with the 3.8 m United Kingdom 
Infrared Telescope (UKIRT) on Mauna Kea, Hawaii. Seeing during the nights was 
FWHM 0\farcs6--0\farcs7. We used the 256$\times$256 pixel IRCAM3 camera, with 
pixel size 0\farcs281 px$^{-1}$ and field of view $\sim70''\times70''$. NGC~
7771 was imaged in the Br$\gamma$ 2.1661~$\mu$m and H$_2$ 1--0 S(1) 
2.122~$\mu$m spectral lines and in the $JHK$ bands. For the emission line 
observations, we used cooled ($T=$ 77 K) narrow-band filters and a 
Fabry-Perot etalon with spectral resolution $\sim$400~km~s$^{-1}$ and 
equivalent width 0.0038~$\mu$m.

For both lines, 5 one-minute observations were made at each on-line 
wavelength (OL) and at the nearby blue (BC) and red (RC) continuum, following 
the sequence OL--BC--OL--RC and moving the telescope by 8$''$ 
between the individual integrations. A dark frame was obtained for each 
sequence. For the $JHK$ bands, 5 one-minute integrations were 
taken both centered on NGC~7771 and on sky 5$'$ east of NGC~7771, moving the 
telescope as in the line observations.

{\em{IRAF}}\footnote{IRAF is distributed by the National Optical Astronomy 
Observatories, which are operated by the Association of Universities for 
Research in Astronomy, Inc., under cooperative agreement with the National 
Science Foundation.}
was used to reduce and analyze the observations. For the line images, a 
normalized flatfield was made from linearized and dark-subtracted galaxy 
images. Galaxy frames were then divided by the flatfield and sky-subtracted 
using the median of the 5 images obtained at the same wavelength. Frames were 
aligned and merged into one on-line image and two continuum images. Since the 
shape of the nucleus varies with wavelength, $JHK$ band images were first 
registered using a foreground field star. The emission line and $K$ band 
images were then registered with respect to each other by using the nucleus 
prior to subtracting the continuum in order to avoid possible rotation and 
alignment errors.  The final registration of the images is accurate to within 
a fraction of a pixel. Finally, the continuum images were scaled with the 
continuum/line ratio measured from observations of a standard star (HD 
225023), combined and subtracted from on-line images, and the final line 
images were then flux calibrated. The total on-line integration time was 40 
minutes for Br$\gamma$ and 20 minutes for S(1). Broad-band images were 
linearized, dark-subtracted, flatfielded using a flatfield made from sky 
images and sky-subtracted using the median of 5 sky images. FS 29 ($K$ = 
13.346) from the Faint Star Catalogue of UKIRT, was used for flux calibration 
and the observed fluxes were corrected for Galactic extinction and redshift 
of NGC~7771 (K--correction).

The wavelength dependence of the Fabry-Perot was corrected for by dividing 
the measured fluxes by an Airy function (see e.g. Bland--Hawthorn 1995). Both 
the velocity field of NGC~7771 and the shift in the transmitted wavelength 
across the array need to be considered. A high resolution velocity field for 
NGC~7771 has not been published, but in a lower resolution data (Nordgren et 
al. 1997) the velocity contours are nearly perpendicular to the major axis of 
the elliptical ring, so the velocity field in the ring was approximated by 
the 1D H$\alpha$ longslit-data (Keel 1996). The effect of linewidth on the 
emission line fluxes has not been corrected for because the shape of the 
lines in the ring is not known. The observed line emission is weak, so in 
order to enhance the S/N, all images were smoothed to 1$''$ resolution. 

\section{Morphology}

The inclined star forming ring with a major axis of $\sim$ 7$''$ (2 kpc at PA 
$\simeq$ 90$^\circ$) is clearly visible in the $K$-band image (Figure 
\ref{fig:7771bg}). The nucleus in the middle of the ring is brighter than the 
circumnuclear regions. In the $H$-band and especially in the $J$-band, the 
morphology is different. The southeastern part of the ring is significantly 
more conspicuous in the $J$-band than in the $K$-band and the nucleus appears 
to merge with the northern part of the ring. In the east the ring is weaker 
than in the west in all the broad-band images. These differences in 
morphology are also clearly visible in the $J$--$H$ color map (Figure 
\ref{fig:7771jh}), where the southern and southwestern parts of the ring are 
redder than other parts, probably due to extinction, RSGs and hot dust.

Although the western part of the ring is weaker in the $K$-band, almost all 
Br$\gamma$ emission arises from there (Figure \ref{fig:7771bg}). The eastern 
part of the ring may be older than the western part and supernova (SN) 
remnants may already have dispersed the gas clouds. Because RSGs appear at a 
later stage in the starburst evolution (Leitherer \& Heckman 1995) and then 
dominate the $K$-band emission, the eastern part may indeed be older. Even in 
the west, the $K$-band peaks do not correlate with the Br$\gamma$ peaks, but 
seem to indicate older starburst regions. There is no Br$\gamma$ emission 
source exactly at the nucleus, but $\sim$0\farcs4 to the north of the nucleus 
there is a strong S(1) source. The position of this strong S(1) peak is 
affected by smoothing. In the unsmoothed image the peak pixel is coincident 
with the nucleus, but emission becomes fainter much faster to the south than 
to the north. This may be due to extended emission or an additional emission 
region north of the nucleus. We believe that the S(1) region marked N is 
really the nuclear emission which has shifted north during the smoothing. The 
position of other emission regions is unaffected by smoothing. The S(1) 
emission in the ring is distributed more broadly than the Br$\gamma$ (Figure 
\ref{fig:7771bg}), especially in the east, delineating a more complete ring 
than what is seen in Br$\gamma$. Neither the correlation between the S(1) and 
$K$-band peaks, nor between the S(1) and Br$\gamma$ peaks is perfect, but 
there is a fair correlation between S(1) and $J$--$H$.

Although the ring is incomplete in Br$\gamma$, it appears to follow an 
ellipse with semi-major and semi-minor axes of $a\simeq$ 3\farcs2 and 
$b\simeq$ 1\farcs2 at PA$\simeq 69^\circ$. This ellipse is different from 
that in the $K$-band ($a\simeq$ 2\farcs8 and $b\simeq$ 1\farcs2 at PA $\simeq 
90^\circ$), but it is consistent with the ellipse in radio emission 
($a\simeq$ 2\farcs75 and $b\simeq$ 1\farcs25 at PA $\simeq 73^\circ$; S99). 
The S(1) emission appears to arise from a ring similar to the Br$\gamma$ 
emission. The $K$-band ellipticity may be different since emission in the 
$K$-band is dominated by RSGs, whereas Br$\gamma$ and radio continuum give 
the orientation of the younger starburst ring which can be affected by gas 
dynamics e.g. in a barred potential. 

We compared our Br$\gamma$ images and the radio images of S99 with after 
registering the Br$\gamma$, radio and K$K$-band images using Fig. 4 of S99. 
There is a reasonable correlation between the Br$\gamma$ and the radio 
emission to the south and southwest of the nucleus (to within 0\farcs2). 
Elsewhere this correlation is worse, as the strongest Br$\gamma$ peaks do not 
have radio counterparts to within a distance of 1$''$. In general, Br$\gamma$ 
correlates better with radio than with the NIR continuum. Morphological 
differences between the Br$\gamma$, radio and NIR continuum images are 
probably due to differences in the ages of the emitting regions, possibly 
indicating several star forming epochs. A similar anticorrelation between the 
radio and NIR continua has also been detected in several other starburst 
systems, e.g. M82 (Golla, Allen \& Kronberg 1996) and NGC~253 (Sams et al. 
1994; Ulvestad \& Antonucci 1997).

\section{Star forming properties}

We detected 9 emission regions defined by the Br$\gamma$ above $3\sigma$ 
level. Furthermore, the region N, located at the nucleus, has no Br$\gamma$ 
but strong S(1) emission. The photometric apertures used were selected to 
avoid overlap between neighboring regions, but to include as much of the 
Br$\gamma$ emission as possible (regions 1--9). The smallest distance between 
the nearest emission regions is 1\farcs2, which is larger than the resolution 
of the smoothed images (1\farcs0), and much larger than the seeing during the 
observations (0\farcs6). Note that there is some overlap between region N and 
region 3. The smallest regions consist of only a few pixels, but these 
regions have radio counterparts (S99) and are therefore probably real. 
Furthermore, as all the regions consist of more than one pixel in unsmoothed 
image, they are unlikely be affected by cosmic rays or other defects. Most of 
the regions are also identifiable when Br$\gamma$ frames were reduced 
separately in two halves. The aperture diameters used, the values of the 
Airy function, the observed and dereddened Br$\gamma$ and S(1) fluxes and the 
$JHK$ colors for the emission regions are given in Table~\ref{tab:7771emi}.

To interpret the star forming properties of NGC~7771, we have used the 
evolutionary model of Leitherer et al. (1999; hereafter L99), which predicts 
various NIR, optical and UV spectral features as a function of metallicity, 
initial mass function (IMF), lower and upper mass cutoff and age, for the 
limiting cases of instantaneous star formation (ISF) and a constant star 
forming rate (CSFR). 

The equivalent width (EW) of hydrogen recombination lines is sensitive to the 
age of the starburst. The Br$\gamma$ EW was estimated by subtracting a smooth 
galaxy model from the $K$-band image and dividing the Br$\gamma$ fluxes with 
the fluxes in the subtracted $K$-band image. The L99 models predict the 
number of ionizing photons, N(H$^0$), which can also be estimated from the 
Br$\gamma$ flux as N(H$^0$) $[s^{-1}]$ = 7.63$\times$ 
10$^{13}$L$_{Br\gamma}$ [erg s$^{-1}$] (Leitherer \& Heckman 1995). The number 
of ionizing photons allows us to evaluate the mass of recently formed stars 
(in ISF) or the star forming rate (in CSFR) via an assumed IMF. N(H$^0$) can 
also be estimated from the thermal radio emission of HII regions (S99): 
N(H$^0$) = 7.1$\times$ 10$^{49}$ D$^2$ $\nu^{0.1}$ T$_e^{-0.76}$ S$_{th}$, 
where D is the distance in Mpc, $\nu$ is the frequency in GHz, $T_e$ is the 
electron temperature in 10$^4$ K and $S_{th}$ is the thermal radio flux 
density in mJy. Finally, the SN rate predicted by the L99 models 
can be compared with that derived from the nonthermal radio emission (Condon 
\& Yin 1990): L$_{NT}$ $\sim$ 1.3$\times$10$^{23}$ $\nu^{-\alpha}$ v$_{SN}$, 
where $L_{NT}$ is the nonthermal radio luminosity in W Hz$^{-1}$, $\nu$ is 
the frequency in GHz, $\alpha$ is the spectral index of nonthermal radiation 
and v$_{SN}$ is the SN rate in yr$^{-1}$. 

Since we do not have enough data for more detailed modeling, we have selected 
two models in what follows: (1) ISF with $M_u$ = 100 M$_\odot$ and $\alpha$ 
= 2.35 and (2) CSFR with $M_u$ = 30 M$_\odot$ and $\alpha$ = 2.35. We have 
assumed solar metallicity. 

The extinction towards the ionized sources was estimated from the
Br$\gamma$/H$\alpha$ ratio. H$\alpha$ imaging has not been published for 
NGC~7771. Veilleux et al. (1995) give an H$\alpha$ flux of 5.9$\times 1
0^{-14}$ ergs s$^{-1}$ cm$^{-2}$ in an aperture of $\sim$ 7$''\times 2''$ (2 
kpc $\times$ 0.55 kpc) that includes most of the ring. We derive A$_V$=2.8 for 
the ring using the Br$\gamma$/H$\alpha$ ratio, Landini et al. (1984) 
extinction law (A$\propto\lambda^{-1.85}$) and a foreground dust screen. 
Similar extinction values were also derived by Davies, Alonso-Herrero \& Ward 
(1997), who suggested an average $A_V$ = 2.3 for the central starbursting 
regions based on the Balmer decrement. If the dust and the emitting gas are 
mixed we derive A$_V$ = 4.0. S99 derived much greater $A_V=5.2$ from the 
H$\alpha$/H$\beta$ ratio, and they suggested an even greater extinction 
($A_V=6.5$) towards the ionized gas from the radio/H$\alpha$ ratio. Higher 
extinction (A$_V>$5 mag) is also suggested by the IR/blue luminosity ratio. 
Extinction probably varies significantly in the ring (see Figure 
\ref{fig:7771jh} for variations in extinction towards the continuum sources), 
but due to the lack of a high resolution H$\alpha$ image, we use the average 
extinction value A$_V$ = 2.8 mag in what follows. 

To provide a check for the extinction, we estimated the extinction towards the 
continuum sources in the ring from the observed $JHK$ colors compared to the 
average $J$--$H$ color of unobscured spiral galaxies ($<$$J$--$H$$>$ = 
0.75, $<$$H$--$K$$>$ = 0.22; Glass \& Moorwood 1985) assuming that only 
extinction affects the colors. The observed $J$--$H$ colors (Figure 
\ref{fig:7771col}) of the ring are redder than the areas inside or outside 
the ring. From the NIR colors we derive a slightly smaller extinction (A$_V$ 
= 0.5--3.9, average A$_V$ = 2.2) than that measured from the recombination 
line ratios, indicating that RSGs responsible for the NIR continuum emission 
are located outside the HII regions surrounding the OB stars. Alternatively, 
this result may also be a selection effect: the most heavily reddened regions 
may not be detectable in our images, and thus we are biased towards the 
low-extinction regions. The southern half of the ring is redder than the 
northern half (Figure \ref{fig:7771jh}), and therefore extinction is probably 
greater in the south, as was noted by Davies et al. (1997). 

The exact value of the extinction does not affect the EW of Br$\gamma$ (ie. 
age) if there is no differential extinction between Br$\gamma$ and continuum 
sources, but it does scale up or down the properties derived from the 
luminosity of Br$\gamma$, including mass, SFR, N(H$^0$) and SN rate. Our 
method of analysis, however, depends on the details of the modeling. Real 
physical starbursts have a finite duration, so the ages derived for ISF are 
really lower limits, as the EW of Br$\gamma$ grows as a function of the burst 
duration for a given age. The observed CO indices (Smith et al. 1996; 
Ridgway, Wynn--Williams \& Becklin 1994) together with the EW of Br$\gamma$ 
can constrain the duration of the burst (Puxley, Doyon \& Ward 1997). The 
exponential decaying time of the burst may have been 1--5 Myr (Puxley et al. 
1997, their Fig. 3), or alternatively $\sim$20 Myr, in which case the burst 
started $\sim$50 Myr ago.
 
The mass of the excited H$_2$ can be estimated by following the procedure of 
Meaburn et al. (1998). The average surface brightness of S(1) in an 8\farcs4 
(2.4 kpc) aperture is 6.4$\times 10^{-13}$ W cm$^{-2}$ sr$^{-1}$ and assuming 
T$_{vib}$ = 2000 K we derive 1740 M$_\odot$ for the mass of the excited H$_2$. 
This value must be multiplied by $\sim$1.1 for the Fabry-Perot calibration 
and by a small, unknown factor for the linewidth, which may be broader 
than the passband width of the Fabry-Perot ($\sim$400 km s$^{-1}$). The 
resulting value $\sim$1900 M$_\odot$ may be compared with the Seyfert 
galaxies NGC~3079 (800 M$_\odot$, Meaburn et al. 1998) and NGC~3227 (250 
M$_\odot$, Fernandez et al. 1999). In all the cases the available mass of HI, 
as derived from the CO emission, is several magnitudes higher. The peak 
surface density of excited hydrogen is 1.9$\times 10^{21}$ m$^{-2}$.

The S(1)/Br$\gamma$ ratio can give clues about the excitation mechanism 
(Puxley et al. 1990). The observed ratio of the emission regions in the 
south and southeast (0.71--0.95) are consistent with UV excitation by young 
stars. Elsewhere in the ring the ratio is slightly larger (1.30--2.10), 
indicating perhaps a relatively larger contribution from shock excitation, 
possibly due to supernovae or stellar winds. Alternatively, gas clouds may be 
denser, or the upper mass cut-off higher in these regions. This seems 
more plausible, because there is no evidence of lower SN rate in the south 
(Table 2; S99). The only emission region outside the ring (4) has the lowest 
value of S(1)/Br$\gamma$ (0.61).

We have detected the nucleus 
in S(1) but not in Br$\gamma$, which is also the case in Seyfert galaxies 
NGC~1097 and NGC~6574 (Kotilainen et al. 1999). But unlike those galaxies, 
the nucleus is not redder than its surroundings. Thus, the extinction in the 
nucleus may be relatively low. The absence of Br$\gamma$ (massive young 
stars) in the nucleus can be explained if the nucleus is either too old to 
harbor OB stars, or the star formation has not yet begun there. 

The EW of H$\alpha$ in the ring is $\sim$10--80 \AA~(Davies et al. 1997), 
which according to the L99 standard model ($\alpha$=2.35, M$_u$ = 100 
M$_\odot$) corresponds to an age range 6.2--11.6 Myr (ISF) or ages $>$1 Gyr 
(CSFR). In the ISF model, the age derived from the EW of Br$\gamma$ (5.5--54 
\AA) is 5.8--6.8 Myr, which would indicate copious numbers of RSGs. Their 
presence within the inner regions of NGC~7771 seems indeed probable because 
of the observed deep NIR CO absorption and the highly enhanced $K$-band 
luminosity (Smith et al. 1996; S99). With the ISF model and using the 
Br$\gamma$ emission, the masses of the star forming regions are 0.3--3.2 
$\times 10^7$ M$_\odot$ (average $1.5 \times 10^7$ M$_\odot$) and SN rates 
are 0.3--3.3 $\times 10^{-2}$ yr$^{-1}$ (average 0.015  yr$^{-1}$). With the 
CSFR model and a reduced upper mass cutoff (M$_u$ = 30 M$_\odot$), as 
suggested by Davies et al. (1997), the ages of several regions remain large, 
$>$ 100 Myr. Such old ages are improbable, as single molecular clouds are 
unlikely to survive more than $\sim$40 Myr (Blitz 1991). The total mass of 
the star forming regions is large, M $\simeq$ 3 $\times10^8$ M$_\odot$, which 
is, however, much less than the total available molecular gas mass derived 
from the CO luminosity, M$_{H_2} \simeq 9\times 10^9$ M$_\odot$ (Sanders, 
Scoville \& Soifer 1991). With the CSFR model, the star forming rate of 
individual regions is 0.2--0.7 M$_\odot$ yr$^{-1}$ (average 0.4 M$_\odot$ 
yr$^{-1}$) and the SN rate predicted by the L99 model is v$_{SN}$ 0.06--1.3 
$\times 10^{-2}$ yr$^{-1}$ (average 0.6$\times 10^{-2}$ yr$^{-1}$). The 
derived star forming properties are given in Table \ref{tab:7771der}. The 
dereddened NIR colors (Figure \ref{fig:7771col}) suggest older ages (8--10 
Myr) than Br$\gamma$ EW. This may be due to the contamination by older 
starburst generations in the NIR continuum after the galaxy model was 
subtracted, which results in artificially low EWs of Br$\gamma$. 
Alternatively, extinction may have been slightly underestimated. 

Our analysis of the star forming properties depends somewhat on the 
evolutionary model which is chosen. L99 employs the newest Geneva 
evolutionary tracks (see e.g. Charbonnel et al. 1999 and references therein) 
and the Lejeune atmosphere models (Lejeune, Buser \& Cuisinier 1997) and as 
such may be considered the most up to date. The general result of a decaying 
starburst remains unchanged if one adopts the evolutionary model of 
Lan\c{c}on \& Rocca-Volmerange (1996). A comparison between the EWs of 
H$\beta$ in L99 and the model of Cervi\~{n}o \& Mas--Hesse (1994) further 
supports the model-independence of our conclusions. However, the {\em{exact}} 
age of the starburst is very model dependent -- one can assume different 
durations or IMFs for the starburst and derive different ages.

The number of ionizing photons derived from the thermal radio emission by S99 
is N(H$^0$) = 5.5$\times 10^{54}$ s$^{-1}$, more than a magnitude larger than 
the N(H$^0$) = 2.0$\times$10$^{53}$ s$^{-1}$ that we derive from Br$\gamma$. 
The simplest explanation for this discrepancy is an underestimation of 
extinction, but this would indicate A$_V \sim 45$ mag, which is unjustified 
by the NIR colors. A difference between the estimates derived from H$\alpha$ 
and the radio continuum was also noticed by S99, but since extinction is much 
higher in the optical wavelengths, they were able to explain the difference 
with $A_V = 6.5$. Alternatively, it is possible that S99 have overestimated 
the fraction of thermal flux in the radio continuum. They suggest that more 
than 50\% of the flux at 6 cm is thermal, at least in the circumnuclear 
region. This is a much larger fraction than what was found in the nuclear 
regions of normal galaxies (between 3\% and 20\%; Turner \& Ho 1994).

An ISF or a short duration model seems more probable than CSFR for NGC~7771, 
based on the clumpy NIR continuum morphology and the observed radio spectral 
index, $\alpha \sim-0.5$ (S99). The L99 model does not contain radio 
continuum information, but the Mas-Hesse \& Kunth (1991) CSFR models cannot 
produce spectral indices steeper than $\sim$--0.3, while their ISF model can.
The ages of the star forming regions derived by us are similar to those 
derived by S99 for the $K$-band regions ($\sim10$ Myr) and the radio regions 
(4--8 Myr), and agree with their suggestion of several star forming 
generations in NGC~7771.

\section{Conclusions}

We have presented the first near--infrared emission line images of the star 
forming ring of NGC~7771, and interpret its star forming properties with the 
L99 evolutionary models. The morphology of NGC~7771 varies strongly with 
wavelength, which may reflect varying ages of star forming regions or several 
stellar generations present in the ring. Also, ellipticities and position 
angles of the rings differ at various wavelengths. No Br$\gamma$ was detected 
in the nucleus, but the nucleus is a strong S(1) emitter. The mass of the 
excited H$_2$ ($M$ $\simeq$ 1900 M$_\odot$) is higher than in the Seyfert 
galaxies NGC~3079 and NGC~3227.

The EW of Br$\gamma$ and the NIR morphology are consistent with an 
instantaneous burst of star formation 6--7 Myr ago. The observed 
Br$\gamma$ EW cannot be produced in the constant star formation model even 
with a lowered upper mass cutoff ($M_u=30$ M$_\odot$) without unlikely large 
ages, greater than 100 Myr for many regions, which are also inconsistent with 
other indicators (e.g. CO absorption indices, SN rates).

\acknowledgements
The United Kingdom Infrared Telescope is operated by the Joint Astronomy 
Centre on behalf of the U.K. Particle Physics and Astronomy Research Council. 
Thanks are due to Tom Geballe and Thor Wold for all the help during the 
observations.
This research has made use of the NASA/IPAC Extragalactic Database (NED), 
which is operated by the Jet Propulsion Laboratory, California Institute of 
Technology, under contract with the National Aeronautics and Space 
Administration.

\newpage 

\figcaption{The $K$-band image of NGC~7771 with Br$\gamma$ (black contours) 
and H$_2$ 1--0 S(1) (white contours) line emission overlaid. The lowest 
contour (3$\sigma$) of Br$\gamma$ is 32\% of the peak ( 5.1$\times 10^{-16}$ 
erg s$^{-1}$ cm$^{-2}$ arcsec$^{-2}$) and other contours are at 50, 70 and 
85\%. In S(1), the contours are at 18, 30, 50, 70 and 85 \% of the peak 
luminosity (7.7$\times 10^{-16}$ erg s$^{-1}$ cm$^{-2}$ arcsec$^{-2}$). The 
arrows point to the Br$\gamma$ regions and to the nucleus mentioned in the 
text.\label{fig:7771bg}}

\figcaption{The $J$--$H$ color map of NGC~7771. The highest contour is 1.0, 
and the contour interval is 0.1\label{fig:7771jh}}

\figcaption{The observed (filled circles) and dereddened (open circles) 
$J$--$H$- and $H$--$K$-colors of NGC~7771. The nucleus has been marked with 
a triangle. The average colors of unobscured spiral galaxies ($<$$J$--$H$$>$ 
= 0.75, $<$$H$--$K$$>$ = 0.22; Glass \& Moorwood 1985) are indicated by an 
ellipse. The average errors of the observed colors, the effects of extinction 
(arrow) and hot dust emission (dotted line) on the colors, and the colors 
predicted by the L99 model (ISF, solid line) are also indicated. The 
tickmarks show the fractional contribution of dust to the $K$-band emission 
at 10 \% intervals.
\label{fig:7771col}}

\begin{deluxetable}{lccccccccccccr}
 \small
 \tablewidth{0pt}
 \tablecaption{Br$\gamma$ emission regions\label{tab:7771emi}}
 \tablehead{
  & & &\multicolumn{4}{c}{observed emission\tablenotemark{a}}&&\multicolumn{5}{c}{dereddened emission\tablenotemark{b}}\\
   \cline{4-7}\cline{9-13}
   \colhead{n}   & \colhead {ap} & \colhead{corr\tablenotemark{c}} & \colhead{Br$\gamma$}    &
   \colhead{S(1)}& \colhead{J--H}& \colhead{H--K} && 
   \colhead{Br$\gamma$} & \colhead{S(1)}& \colhead{$\frac{S(1)}{Br\gamma}$} & 
   \colhead{J--H}& \colhead{H--K}& \colhead{A$_V$\tablenotemark{d}}         \\
   & \colhead{$''$}& &\multicolumn{2}{c}{10$^{-15}$ ergs s$^{-1}$ cm$^{-2}$}
   & \colhead{mag} 
   & \colhead{mag} & &\multicolumn{2}{c}{10$^{-15}$ ergs s$^{-1}$ cm$^{-2}$} 
  && \colhead{mag} &   \colhead{mag}&   \colhead{mag}
 }
\startdata
 1 & 2.2& 1.52& 0.59& 1.23& 0.84& 0.33&& 0.73& 1.54& 2.10 & 0.59 & 0.18& 1.1 \\
 2 & 1.4& 1.11& 0.70& 1.44& 0.82& 0.36&& 0.87& 1.80& 2.06 & 0.57 & 0.21& 1.0 \\
 3 & 2.2& 1.03& 0.91& 1.76& 0.91& 0.42&& 1.13& 2.20& 1.94 & 0.66 & 0.27& 2.1 \\
 4 & 1.7& 1.02& 0.35& 0.21& 0.78& 0.18&& 0.44& 0.26& 0.61 & 0.53 & 0.03& 0.5 \\
 5 & 1.4& 1.01& 0.41& 0.54& 0.92& 0.42&& 0.51& 0.68& 1.32 & 0.67 & 0.26& 2.2 \\
 6 & 1.0& 1.01& 0.27& 0.24& 1.01& 0.52&& 0.33& 0.29& 0.89 & 0.76 & 0.37& 3.3 \\
 7 & 1.4& 1.02& 0.62& 0.58& 1.06& 0.58&& 0.77& 0.73& 0.95 & 0.81 & 0.42& 3.9 \\
 8 & 1.4& 1.02& 0.45& 0.34& 1.00& 0.53&& 0.56& 0.43& 0.76 & 0.75 & 0.37& 3.1 \\
 9 & 1.4& 1.06& 0.46& 0.32& 0.93& 0.47&& 0.57& 0.40& 0.71 & 0.67 & 0.32& 2.2 \\
 N & 1.1& 1.01& --  & 2.61& 0.98& 0.46&& --  & 3.24& --   & 0.73 & 0.31& 2.9 \\
\enddata
\tablenotetext{a}{The $JHK$ magnitudes and colors are accurate to within 0.03 
and 0.05 mag, respectively. Our observed colors agree well with those 
reported by S99. For the narrowband photometry, accuracy varies from 8\% 
(region 3) to 12 \% (region 1), with average value is 9\%.}
\tablenotetext{b}{Dereddened using  A$_V=2.8$}
\tablenotetext{c}{Correction factor for the velocity field and the change of wavelength across the array}
\tablenotetext{d}{Obtained from the NIR color: $<$$J$--$H$$>$ = 0.75 (Glass \& Moorwood 1985) }
\end{deluxetable}

\begin{deluxetable}{lccccccccccr}
 \small
 \tablewidth{0pt}
 \tablecaption{Star forming properties\label{tab:7771der}}
 \tablehead{ 
  & & &  \multicolumn{4}{c}{instant star formation}  &
         \multicolumn{4}{c}{constant star forming rate}\\
  \cline{4-7} \cline{9-12}
  \colhead{n}    & \colhead{N(H$^0$)}& \colhead{EW}      & \colhead{age}     &
  \colhead{mass} & \colhead{SFR\tablenotemark{a}}        & \colhead{v$_{SN}$}&&
  \colhead{10$^6$ yr}  & \colhead{SFR}     & \colhead{mass\tablenotemark{b}}       &
  \colhead{v$_{SN}$}                 \\
                 &  \colhead{10$^{52}$ s$^{-1}$} &  \colhead{\AA}  &  
  \colhead{10$^6$ yr}  & \colhead{10$^{6}$ M$_\odot$} & \colhead{M$_\odot$ y$^{-1}$} &  \colhead{10$^{-3}$ yr$^{-1}$}  && 
  \colhead{Myr}  & \colhead{M$_\odot$ y$^{-1}$} & \colhead{10$^{6}$ M$_\odot$} &  \colhead{10$^{-3}$ yr$^{-1}$} 
}
\startdata
  1& 2.14 & 18.7\phn & 6.37 &    12.6  & 1.98  &   11.9&& \phn11.8    & 0.46 & \phn\phn5 &\phn2.3\\
  2& 2.55 & \phn8.55 & 6.63 &    20.0  & 3.01  &   18.4&& \phn48.2    & 0.55 & \phn26    &\phn9.2\\
  3& 3.32 & \phn5.46 & 6.81 &    32.2  & 4.73  &   32.6&& 176\phd\phn & 0.71 &    125    &   12.5\\
  4& 1.28 & 54.0\phn & 5.84 & \phn3.4  & 0.59  &\phn2.9&& \phn\phn8.4 & 0.28 & \phn\phn2 &\phn0.6\\
  5& 1.50 & \phn6.58 & 6.72 &    13.6  & 2.02  &   12.5&& 100\phd\phn & 0.32 & \phn32    &\phn6.7\\
  6& 0.97 & \phn7.93 & 6.65 & \phn7.8  & 1.18  &\phn7.2&& \phn59.5    & 0.21 & \phn12    &\phn3.9\\
  7& 2.26 & \phn8.91 & 6.62 &    17.4  & 2.63  &   16.1&& \phn43,6    & 0.49 & \phn21    &\phn9.2\\
  8& 1.65 & \phn6.50 & 6.72 &    14.9  & 2.22  &   13.8&& 104\phd\phn & 0.35 & \phn37    &\phn6.2\\
  9& 1.66 & \phn5.76 & 6.78 &    15.8  & 2.34  &   15.5&& 149\phd\phn & 0.36 & \phn53    &\phn6.9\\
\enddata
\tablenotetext{a}{Mass divided by age}
\tablenotetext{b}{SFR multiplied by age}
\end{deluxetable}

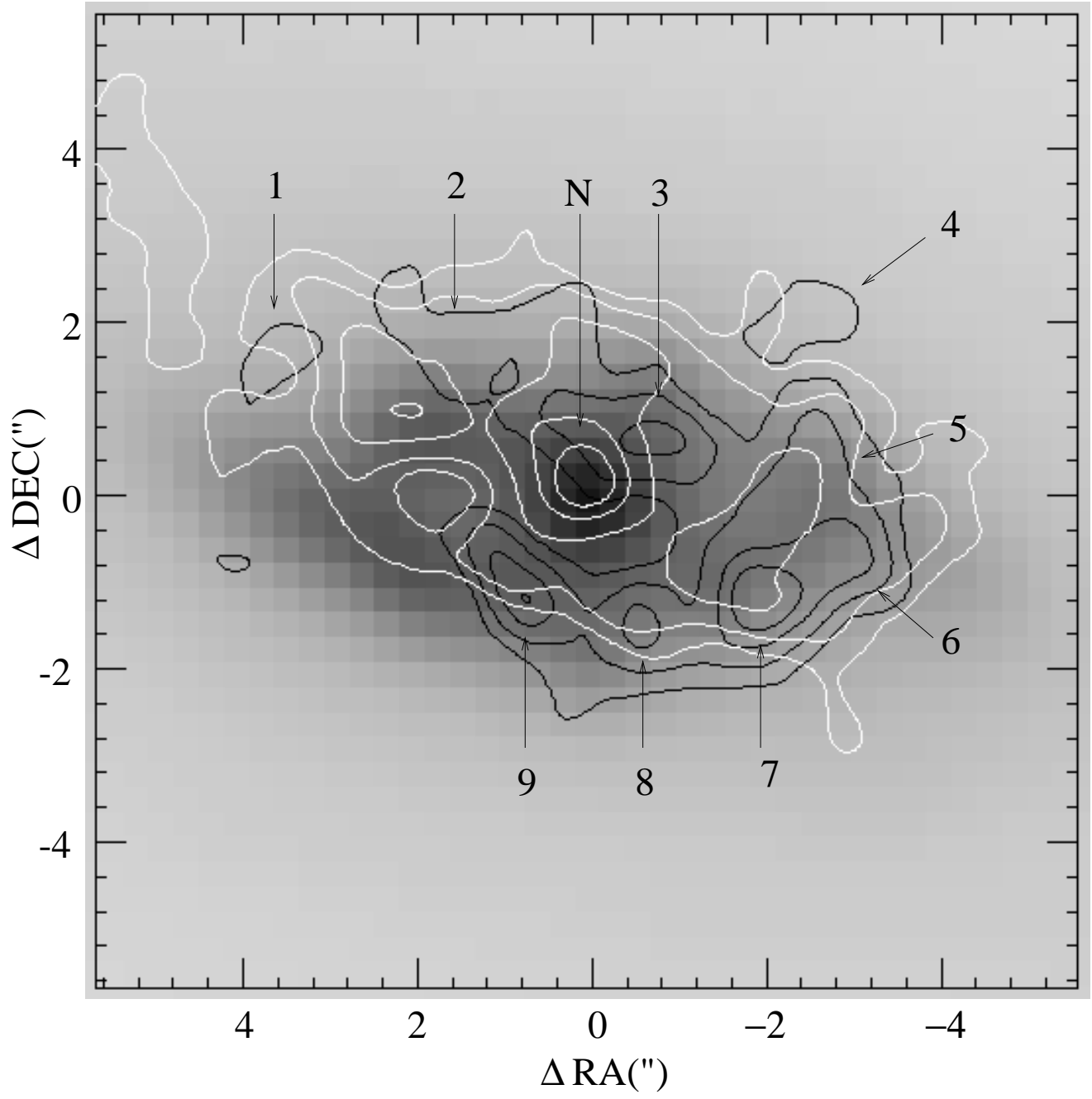
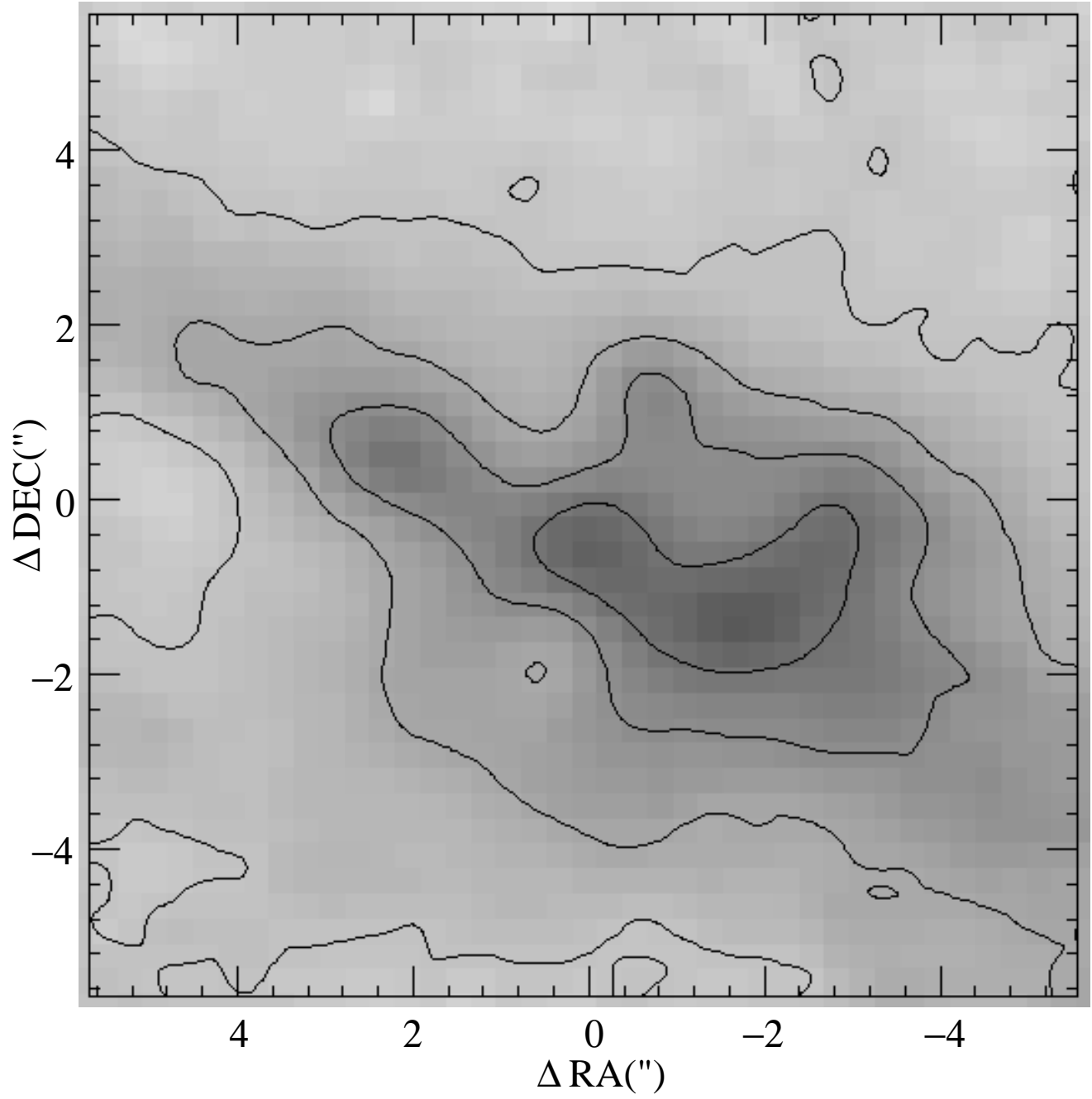
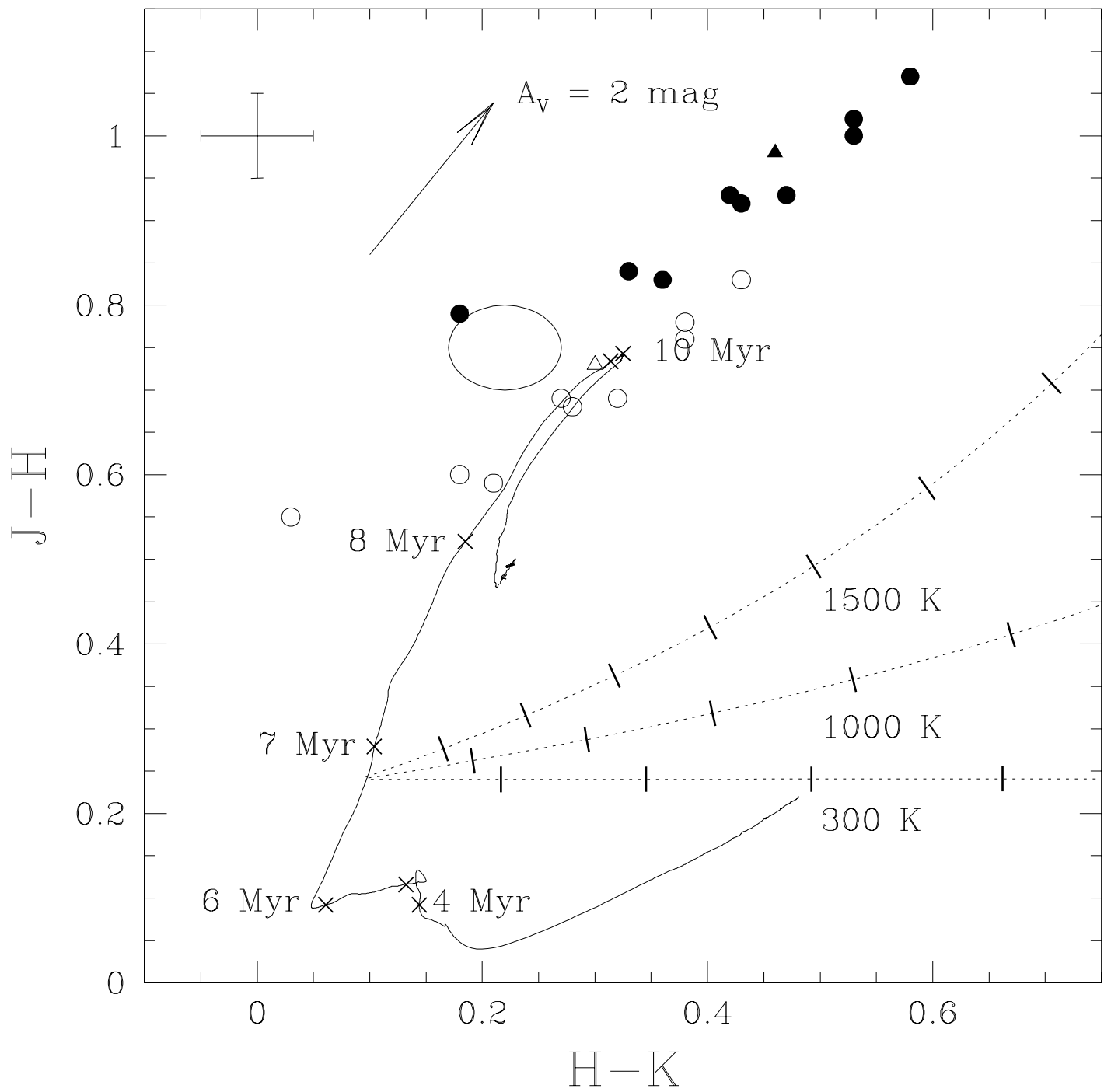
\end{document}